\documentclass[sigconf]{acmart}

\usepackage{multirow}
\usepackage{xspace}
\usepackage{colortbl,array}

\AtBeginDocument{%
  }

\copyrightyear{2025}
\acmYear{2025}
\setcopyright{cc}
\setcctype{by}
\acmConference[HAI '25]{Proceedings of the 13th International Conference on Human-Agent Interaction}{November 10--13, 2025}{Yokohama, Japan}
\acmBooktitle{Proceedings of the 13th International Conference on Human-Agent Interaction (HAI '25), November 10--13, 2025, Yokohama, Japan}
\acmPrice{}
\acmDOI{10.1145/3765766.3765767}
\acmISBN{979-8-4007-2178-6/25/11}

\usepackage{enumitem}
\setenumerate[1]{itemsep=0pt,partopsep=0pt,parsep=0pt,topsep=0pt,leftmargin=0pt}
\setitemize[1]{itemsep=0pt,partopsep=0pt,parsep=0pt,topsep=0pt,leftmargin=12pt}
\setdescription{itemsep=0pt,partopsep=0pt,parsep=0pt,topsep=0pt,leftmargin=0pt}

\begin{document}

\newcommand{\Gui}[1]{{\color{green}#1}}
\newcommand{\Li}[1]{{\color{cyan}#1}}

\newcommand{\Scenario}{\textit{Scenario}\xspace}
\newcommand{\Emotion}{\textit{Emotion-Communication}\xspace}

\newcommand{\Trust}{\textit{Trust}\xspace}
\newcommand{\Understanding}{\textit{Understanding}\xspace}


\newcommand{\iqr}[2]{$IQR={#1}-{#2}$}
\newcommand{\m}{\textit{M=}}

\newcommand{\md}{\textit{Mdn=}}
\newcommand{\sd}{\textit{SD=}}

\newcommand{\N}{\textit{N=}}

\newcommand{\padjminor}{\textit{p$_{adj}<$}}
\newcommand{\padj}{\textit{p$_{adj}$=}}

\newcommand{\ttest}{\textit{t=}}
\newcommand{\ns}{\textit{ns.}}
\newcommand{\V}[1]{$V={#1}$}
\newcommand{\F}[3]{$F({#1},{#2})={#3}$}
\newcommand{\W}{\textit{W=}}
\newcommand{\Z}{\textit{Z=}}
\newcommand{\p}{\textit{p=}}

\newcommand{\rtest}{\textit{r=}}

\newcommand{\df}{\textit{df=}}
\newcommand{\pminor}{\textit{p$<$}}
\newcommand{\kendall}{\textit{r}$_\tau$=}
\newcommand{\chisq}{$\chi^2$}
\newcommand{\effectsize}{\textit{r=}}
\newcommand{\rsq}{$\textit{R}^2$=}
\newcommand{\baseline}{\textit{baseline}\xspace}
\newcommand{\coef}{$\beta$=}

\title{TailCue: Exploring Animal-inspired Robotic Tail for Automated Vehicles Interaction}

\author{Yuan Li}
\authornote{Both authors contributed equally to this research.}
\orcid{0009-0005-8013-0896}
\affiliation{%
  \institution{The University of Tokyo}
  \state{Tokyo}
  \country{Japan}
}
\email{liyuan@keio.jp}

\author{Xinyue Gui}
\orcid{0000-0001-6541-224X}
\authornotemark[1]
\affiliation{%
  \institution{The University of Tokyo}
  \city{Tokyo}
  \country{Japan}
}
\email{xinyueguikwei@gmail.com}

\author{Ding Xia}
\orcid{0000-0002-4800-1112}
\affiliation{%
  \institution{The University of Tokyo}
  \city{Tokyo}
  \country{Japan}
}
\email{dingxia1995@gmail.com}

\author{Mark Colley}
\orcid{0000-0001-5207-5029}
\affiliation{%
  \institution{UCL Interaction Centre}
  \city{London}
  \country{United Kingdom}
}
\email{m.colley@ucl.ac.uk}

\author{Takeo Igarashi}
\orcid{0000-0002-5495-6441}
\affiliation{%
  \institution{The University of Tokyo}
  \city{Tokyo}
  \country{Japan}}
\email{takeo@acm.org}

\renewcommand{\shortauthors}{Li, Gui et al.}

\begin{abstract}
Automated vehicles (AVs) are gradually becoming part of our daily lives. However, effective communication between road users and AVs remains a significant challenge. Although various external human-machine interfaces (eHMIs) have been developed to facilitate interactions, psychological factors, such as a lack of trust and inadequate emotional signaling, may still deter users from confidently engaging with AVs in certain contexts. To address this gap, we propose TailCue, an exploration of how tail-based eHMIs affect user interaction with AVs. We first investigated mappings between tail movements and emotional expressions from robotics and zoology, and accordingly developed a motion-emotion mapping scheme. A physical robotic tail was implemented, and specific tail motions were designed based on our scheme. An online, video-based user study with 21 participants was conducted. Our findings suggest that, although the intended emotions conveyed by the tail were not consistently recognized, open-ended feedback indicated
that the tail motion needs to align with the scenarios and cues. Our result highlights the necessity of scenario-specific optimization to enhance tail-based eHMIs. Future work will refine tail movement strategies to maximize their effectiveness across diverse interaction contexts.
\end{abstract}



\begin{teaserfigure}
  \centering
  \includegraphics[width=1\linewidth]{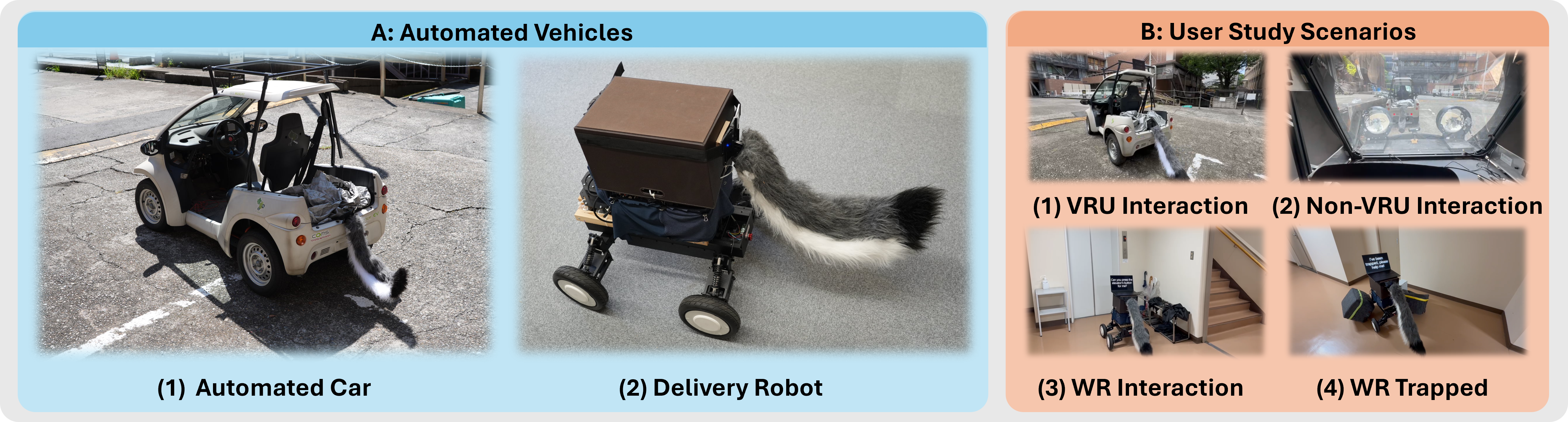}
  \caption{Overview of this study. (A) The two types of automated vehicles, a car and a delivery robot, both equipped with physical tail-based eHMI, were implemented in this study. (B) illustrate the four AV interaction scenarios for our user study. (B) - 1 and 2 demonstrate the scenario interacted with an automated car as vulnerable and non-vulnerable road users. (B) - 3 and 4 demonstrate the 2 scenarios for delivery robot interaction. Specifically, (B) - 1 scenario of drivers changing lanes.  (B) - 2 Scenario of car-pedestrian interaction. (B) - 3 Weak Robot (WR) seeking the help of pedestrians to press the elevator button. (B) - 4  WR seeking help to remove the object that trapped it. More details regarding the scenario will be provided in the user study section.}
  \label{fig:teaser}
\end{teaserfigure}

\maketitle

\section{Introduction}

Automated vehicles (AVs) are one instance of an agent on the road that is increasingly integrated into our urban environments. As AVs begin to operate without human drivers, they require effective communication with other road users, especially pedestrians, cyclists, and other vulnerable road users (VRUs)~\cite{markVulnerable}. For instance, when a pedestrian attempts to cross a street without a traffic signal, clear communication from the AV is essential. In such scenarios, interfaces are necessary to convey the vehicle’s status, intention, or awareness.

External Human-Machine Interfaces (eHMIs) have been developed to address this challenge, such as flashing LED lights, roadway projections, and embedded displays~\cite{GuiArmeHMI, help-seekingeHMIdesign, guigazing, IntenteHMIscreen, 10.1145/3544548.3581303, 10.1145/3546717, COLLEY2022120, 10.1145/3596248, COLLEY2022303, 10.1145/3706598.3714187, 10.1145/3613904.3642031, 10.1145/3699778, 10.1145/3546711, 10.1145/3447526.3472024}. These solutions have shown promising results in improving pedestrians’ understanding of AV behavior. However, some pedestrians may still hesitate to interact with the AVs. For example, when crossing the street, pedestrians may understand the intention of the AVs through their eHMI, yet still hesitate to cross due to psychological concerns. This may be caused by the fact that the interaction between pedestrians and vehicles is often bidirectional. Pedestrians typically wait for confirming cues from the vehicle before confirming with their own crossing behavior~\cite{eHMIsaferhuman, eHMIemojidecisionquickly}. Pedestrians require a trustworthy signal before they are willing to respond or take action~\cite{10.1145/3409118.3475133}. 
Similarly, in the case of smaller AVs such as delivery robots, when a temporary malfunction occurs, nearby pedestrians may understand the situation but remain unwilling to offer assistance~\cite{hurtrobot->help}. These cases indicate that simply conveying information or intent may not be enough. Therefore, it is also essential to find a more acceptable approach that encourages road users to accept and engage with AVs. 

Existing studies have attempted to use gamification to encourage user interaction with AVs~\cite{GamificationeHMI}. However, gamified approaches may pose potential safety risks in complex road environments. It is necessary to identify more reliable methods for fostering user engagement. In Human-robot interaction (HRI), the community has used different external interfaces to enhance interaction experiences. Animal-inspired interfaces, including tails, eyes, ears, etc., have proven effective in encouraging human interaction, as animals have historically formed emotional bonds with humans~\cite{animalHRI, animalHRIAibogood, Dogcommunication, HRIPelo}. Likewise, in zoology, we have similar discoveries~\cite{ ZOOCattailandearDeputte, ZOOcatunderstandingbehaviorEllis}. Animals use their bodily cues and visual expressions to build bonds and trust with human beings.  Especially for tails, which are one of the most prominent parts of an animal’s body, with socially meaningful parts of many animals~\cite{ZOOallanimalKiley}. Compared to features such as eyes or ears, the tail is more suitable as an interface for outdoor scenarios due to its higher visibility and expressiveness, and the tail can trigger users' empathy as a pet-like robot~\cite{qoobo, ROBOsinghDogtailfullimp}. These results suggest that implementing tail as an eHMI may offer a more natural and trust-enhancing channel for communication in the context of AV interactions. However, deploying tail-based eHMI in AV scenarios presents several challenges. AV interactions typically occur in outdoor public environments, which are far more complex and unpredictable than controlled indoor settings, such as those pet robots mentioned above. Users must process a multitude of environmental cues, increasing their cognitive load and making the effectiveness of tail-based communication in these contexts uncertain~\cite{eHMIsurvey, eHMIsaferhuman}. This makes introducing tail into the AV interaction scenario even more complex compared to the traditional HRI scenario. It remains unclear whether tail-based communication is effective under such conditions.
To address this, we raised the research question (RQ): 
\begin{quote}
\textit{RQ: What are the effects of a tail-based eHMI on interactions between AVs and pedestrians?} 
\end{quote}

To answer our RQ, we first developed a mapping scheme, translating a set of tail motions to convey emotions informed by previous research on tail behaviors and emotional expression in both zoology and HRI. 
We implemented a physical prototype by modifying a commercial robotic tail, mounted on an AV and delivery robot (\autoref{fig:teaser}), and implemented our designed tail movements based on our proposed mapping scheme and tail function, so that the vehicle can drive with the tail moving. After this, we then defined four scenarios (two for AVs, two for delivery robots) and conducted a user study including N=21 participants. Our results indicate that in the context of AV interactions, the emotion expressed by the tail is hardly recognized. Open feedback results suggest that the tail motion needs to align with the scenario and cues. Scenario-specific optimization is required for better interaction between AVs and road users. Future research will focus on refining tail movement strategies to maximize effectiveness across different interaction contexts. 

\smallskip

\textit{Contribution Statement:} 
\begin{itemize}
    \item We propose and implement a set of tail motions, including the mapping between specific tail movements and emotional expressions.
    \item We conducted a user study including 21 participants to investigate the effectiveness of tail-based eHMI in automated vehicle interaction scenarios.
\end{itemize}

\section{Related Work}

\subsection{External Interface for Road Interaction}



External Human-Machine Interfaces (eHMIs) can inform pedestrians about the intentions of automated cars when there is no driver. Various types of eHMI have already been developed, and they can be classified into 2 communication types: linguistic and extra-linguistic~\cite{Linguistconcept, Linguisticexpression}. Linguistic communication encompasses spoken or written language, while extra-linguistic communication refers to forms of communication that occur beyond spoken or written language, including bodily cues, visual expressions, and non-speech sounds. Each employs different methods to convey intent to road users to address current challenges.

Linguistic communication is the most straightforward to read and understand, as it uses explicit language. However, they are not always well accepted by all users. One approach utilizes linguistic cues~\cite{Linguistconcept} to directly communicate the vehicle’s intention to pedestrians, typically using text displays on windows or additional screens mounted on the vehicle~\cite{gui2024text+, eHMIsurvey}. However, while easy to comprehend, this approach can be limited by cultural differences and environmental lighting conditions. For example, children may have difficulty understanding text, and standard displays may not be sufficiently visible in bright sunlight~\cite{eHMIsurvey}. Furthermore, explicit language alone may not be sufficient to persuade pedestrians to trust AVs, as a significant portion of the public remains skeptical of AI-driven service machines~\cite{eHMIsurvey, ROBOtailsurveysaab}. The second approach involves extra-linguistic communication~\cite{Linguisticexpression}, such as using human-like arms to gesture to pedestrians~\cite{GuiArmeHMI} or employing eye-based interfaces that simulate eye contact~\cite{guigazing}. These methods guide road users through intuitive, observable actions, but often lack the visibility~\cite{guigazing} in far distance, or lack the ability to convey emotion~\cite{GamificationeHMI}. On the other hand, tail-based eHMI has shown considerable effectiveness in terms of user acceptance and emotional engagement within pet robot contexts~\cite{qoobo, ROBOcatliketail}. To the best of our knowledge, the element of emotional expression still remains unexplored in the context of AV interaction scenarios. This motivated us to explore a method that could express emotion to encourage interaction between AVs and pedestrians. Many animal-inspired external interfaces have been shown to effectively convey emotions. For example, ears, wings, tails~\cite{animalHRIAibogood, ROBOcatliketail, ROBObioinspconveyemotion}. Among these, we selected the tail, as it is one of the most visible parts of an animal’s body and is particularly well-suited for outdoor environments, allowing pedestrians to clearly observe its movements even from a distance.


\subsection{Tail Applications in Life Science and Interaction Technology}

The tail is a significant anatomical feature in many animal species. 
Two widely recognized functions of animal tails are: (1) assisting body balance in locomotion and (2) conveying emotions and physiological states~\cite{ZOOallanimalKiley}. We focus on the latter one in this study. In HRI, studies have explored how tails can be implemented on robots to facilitate emotional expression and enhance robot-human communication. Saab et al. reviewed the structures for existing implemented robotic tails and classified them into four categories~\cite{ROBOtailsurveysaab}: Single-body rigid, Spatial, articulated, and continuum. Among these, types articulated, and continuum are most commonly employed for HRI scenarios~\cite{ROBOxietail,ROBOtaildesign, ROBOsinghDogtailfullimp, ROBOsinghDogTailIntention, ROBOcatliketail}. The primary purpose of robotic tails is to make robots appear more approachable and to enhance user acceptance by incorporating tail movements. To do this, robotic tails require considerations of mechanical feasibility and intuitive motion for human users to correctly interpret the robot’s emotional expressions and respond appropriately. Currently, the use of mechanical tails on pet robots has been shown to encourage human-robot interaction and evoke empathy in users~\cite{ROBOsinghDogtailfullimp, ROBOcatliketail, ROBOsinghDogTailIntention, ROBOismarVRtail}. These features inspired us to introduce a mechanical tail as an eHMI in AV interaction scenarios. However, there are substantial differences between AV interactions and those involving indoor pet robots. Pet robots typically interact with human users in the stable and predictable environment of the home. In contrast, AV interaction scenarios take place outdoors, in shared public spaces, where both users and environmental conditions are less predictable. This major scenario gap motivated our study.


\begin{table*}[ht!]
\centering
\footnotesize
\setlength{\tabcolsep}{12pt}
\caption{Tails in the animal domain: which motion elements convey different types of emotion by animal species.}
\label{tab:emotions}
\begin{tabular}{lllll}
\toprule
\multicolumn{2}{l}{\begin{tabular}[c]{@{}l@{}} \textbf{Emotion}\\ (Ekman~\cite{ekman1999basic})\end{tabular}} &
  \textbf{Tail Behavior} &
  \textbf{Species} &
  \textbf{Ref.} \\ \midrule
\multicolumn{2}{l}{\multirow{2}{*}{Happy}} &
  Tail raised, wagging continuously &
  Pig &
 \cite{ZOOPigtailemotionMarcet} \\
\multicolumn{2}{l}{} &
  Tail raised, high wag amplitude &
  Dog &
 \cite{ZOOdogspecialRuge} \\ \midrule
\multicolumn{2}{l}{\multirow{3}{*}{\begin{tabular}[c]{@{}l@{}}Social Approach\\ (Surprised)\end{tabular}}} &
  Tail upright, vertical waggling vertical &
  Cat &
 \cite{ZOOcatunderstandingbehaviorEllis, ZooCatzhao} \\
\multicolumn{2}{l}{} &
  Tail upright, high wag amplitude &
  Dog &
 \cite{ZoodogtaildirectionSiniscalchi} \\
\multicolumn{2}{l}{} &
  Curled tail, continuous waggling &
  Pig &
 \cite{ZOOPigreviewCamerlink} \\ \midrule
\multicolumn{2}{l}{\begin{tabular}[c]{@{}l@{}}Curiosity\\ (Surprised)\end{tabular}} &
  Tail lightly raised or curled &
  Cat &
 \cite{ZooCatzhao} \\ \midrule
\multicolumn{2}{l}{\multirow{2}{*}{\begin{tabular}[c]{@{}l@{}}Fear \\ (Scared)\end{tabular}}} &
  Tail hanging down, tucked tightly against the body &
  Pig &
 \cite{ZOOPigtailemotionMarcet, ZOOPigreviewCamerlink} \\
\multicolumn{2}{l}{} &
  Tail curved, hidden under body &
  Cat &
 \cite{ZOOcatunderstandingbehaviorEllis} \\ \midrule
\multicolumn{2}{l}{\multirow{2}{*}{\begin{tabular}[c]{@{}l@{}}Frustration\\  (Sad)\end{tabular}}} &
  Tail flicking rapidly &
  Cat &
 \cite{ZooCatzhao} \\
\multicolumn{2}{l}{} &
  Still or Low tail with low waggling amplitude &
  Dog &
 \cite{ZOOdogspecialRuge} \\ \midrule
\multicolumn{2}{l}{\multirow{2}{*}{\begin{tabular}[c]{@{}l@{}}Defensiveness\\  (Angry, Disgust)\end{tabular}}} &
  Low tail, left-biased wag &
  Dog &~\cite{ZoodogtaildirectionSiniscalchi} \\
\multicolumn{2}{l}{} &
  Tail wrapped close to the body &
  Cat &
 \cite{ZOOcatunderstandingbehaviorEllis} \\ \bottomrule
\end{tabular}
\end{table*}

\begin{table*}[ht]
\setlength{\tabcolsep}{4pt}
\renewcommand{\arraystretch}{1.1}
\footnotesize
\centering
\caption{Summary of tail research in HRI: the relationship between robotic control parameters, tail modalities, and the different applications based on the basic Ekman emotion model.}
\label{tab:hri}
\resizebox{\textwidth}{!}{%
\begin{tabular}{@{}lllll@{}}
\toprule
\textbf{Ref.} &
  \textbf{Control Parameters} &
  \textbf{Output Modalities} &
  \textbf{Emotion Model} &
  \textbf{Emotion Mapping} \\ \midrule
\cite{ROBOsinghDogtailfullimp, ROBOsinghDogTailIntention} &
  \begin{tabular}[c]{@{}l@{}}Speed, Wag-size, Height, \\ Wagging type (Horizontal, Vertical, Circular), \\ Postures (high, low, neutral)\end{tabular} &
  Physical Tail &
  Circumplex &
  \begin{tabular}[c]{@{}l@{}}Higher speed → Higher arousal/valence; \\ Smaller wag-size → More energetic; \\ Higher tail → Happier; \\ Horizontal wag → Positive valence; \\ Vertical wag → Negative valence/aggression; \\ Circular wag → Mixed but generally high arousal\end{tabular} \\ \midrule 
\cite{ROBOtailLeeStudleyIntention} &
  Height, Speed, Wag (side-to-side intensity) &
   &
  Custom &
  N/A \\ \midrule
\cite{ROBOcatliketail} &
  Frequency, Amplitude, Waveform &
  Physical Tail + Eyes &
  Modified Ekman &
  \begin{tabular}[c]{@{}l@{}}High amplitude/frequency → Enhances Happy/Angry; \\ Low amplitude/frequency → Enhances Sadness; \\ Square wave → Enhances Anger specifically\end{tabular} \\ \midrule
\cite{ROBOghafurianMiro} &
  \begin{tabular}[c]{@{}l@{}}Position (Up, Halfway up, Down), \\ Wagging type (Still, Horizontal, Verticle), \\ Speed (Slowly, Fast)\end{tabular} &
  \begin{tabular}[c]{@{}l@{}}Muti-Modal \\ (eyes, neck, ear and etc.)\end{tabular} &
  Custom &
  \begin{tabular}[c]{@{}l@{}}Up, wagging horizontal widely → Happy; \\ Down, Still → Sad; \\ Up, Wagging Horizontal, widely, fast → Excited; \\ Wagging Vertically → Fearful; \\ Up, Still → Disgusted / Surprised; \\ Halfway up, slowly, waggling horizontally → Calm; \\ Down, slowly, Waggling Horizontally → Bored; \\ Wagging Vertically, slowly → Annoyed; \\ Up, still → Angry; \\ Down, still → Tired\end{tabular} \\ \midrule
\cite{ROBOismarVRtail} &
  \begin{tabular}[c]{@{}l@{}}Pitch (Width), \\ Yaw (Raised, Lowered), \\ Rotational Speed (6, 20, 100 degree), \\ Range (Vertical, Vertical, Horizontal)\end{tabular} &
  VR Tail &
  Modified Ekman &
  \begin{tabular}[c]{@{}l@{}}Width(70), Raised (15) Speed:20/s, Horizontal Wag → Happy; \\ Width(28), Lowered(10) Speed:6/s, Vertical Wag → Sad;\\ Width(10), Raised(25) Speed:100/s, Vertical → Angry;\\ Width(2), Lowered(20) Speed:100/s, Horizontal Wag → Scared\\ Width(4), Raised(35) Speed:100/s, Horizontal Wag → Surprised\end{tabular} \\ \bottomrule
\end{tabular}%
}
\end{table*}

\newcommand{\na}{--}
\begin{table*}[ht]
\centering
\footnotesize
\setlength{\tabcolsep}{6pt}
\renewcommand{\arraystretch}{1.2}
\caption{The mapping scheme decoding the emotion (from Ekman's 6 motion model) with tail commonly accepted and practically implementable motion elements extracted from tables 1 and 2.}
\label{tab:my-table}
\begin{tabular}{l|ccc|cc|c|cc|cc|c}
\toprule
\multirow{4}{*}{\raisebox{-2.5ex}{\textbf{Emotion}}}
    & \multicolumn{11}{c}{\textbf{Tail motion}} \\ 
    \cmidrule{2-12}
& \multicolumn{6}{>{\columncolor{blue!12}}c|}{\textit{Continuous Wagging}} 
    & \multicolumn{4}{>{\columncolor{orange!12}}c|}{\textit{Action Gestures}} 
    & \multicolumn{1}{>{\columncolor{green!12}}c}{\textit{Static Postures}} \\ 
    \cmidrule(lr){2-7} \cmidrule(lr){8-11} \cmidrule(l){12-12}
& \multicolumn{3}{c|}{Horizontal} & 
    \multicolumn{2}{c|}{Vertical} & 
    Circular & 
    \multicolumn{2}{c|}{Raising} & 
    \multicolumn{2}{c|}{Lowering} & 
    \multirow{2}{*}{\raisebox{-0.8ex}{Height}}  \\
    \cmidrule(lr){2-4} \cmidrule(lr){5-6} \cmidrule(lr){7-7} \cmidrule(lr){8-9} \cmidrule(lr){10-11} 
& Speed & Wag-Size & Height 
    & Speed & Wag-Size 
    & Speed 
    & Speed & Height
    & Speed & Height &  \\
    \midrule
Happy    & High & Large & High & \na & \na & High & \na & \na & \na & \na & High \\
Angry    & \na  & \na   & \na  & High & Large & Medium & \na & \na & \na & \na & \na \\
Sad      & \na  & \na   & \na  & \na & \na & \na & \na & \na & Low & Low & \na \\
Scared   & High & Small   & Low  & \na & \na & \na & \na & \na & \na & \na & Medium \\
Suprised & \na  & \na   & \na  & \na & \na & \na & Fast & High & \na & \na & \na \\
Disgust  & Medium & Large & High & Medium & Large & \na & Medium & High & \na & \na & \na \\
\bottomrule
\end{tabular}
\end{table*}

\section{Mapping Scheme}

This section describes the mapping scheme development for the relationship between tail movements and emotional expression, drawing on previous research in zoology and HRI. Based on the zoological literature, we summarized the functional roles of animal tails and the method by which animals use tail movements to communicate intent with emotions. By reviewing previous studies that used tails as external interfaces, we identified how specific tail movements correspond to emotional expressions that humans can intuitively understand. This helps us to generate motions with human-interpretable emotions, and adapts the concept of Ekman's 6 basic emotion model for later design~\cite{ekman1999basic}. This mapping scheme served as a reference for our physical tail prototype.  


\subsection{Contradictions in Tail Behavior Interpretation}
We identified various previous works that often report contradictory results in zoology and HRI. For example, rapid tail flicking is interpreted by some researchers as a sign of internal conflict or anxiety~\cite{ZooCatzhao}, especially when the cat appears indecisive or hesitant. In contrast, others associate this same behavior with pre-aggressive states~\cite{ZOOcatunderstandingbehaviorEllis}. Similarly, gentle tail swaying is sometimes described as a signal of relaxation~\cite{ZooCatzhao}, while earlier behavioral studies suggest it could be a displacement activity triggered by mild stress or environmental uncertainty~\cite{ZOOallanimalKiley}. Based on this exploration, we first extracted commonly accepted, widely used, and practically implementable elements. This allows us to establish a clear guideline for our evaluation and provides a useful foundation for other researchers conducting similar research in the future.

\subsection{Tail in Animal Domain - \autoref{tab:emotions}} 

We began by exploring research in zoology on the mapping between tail movements and emotions. Our exploration covers a wide range of mammals, including cats, dogs, pigs, sheep, horses~\cite{ZooCatzhao, ZOOallanimalKiley, ZOOdogspecialRuge, ZOOPigreviewCamerlink, ZOOCattailandearDeputte, ZOOPigtailemotionMarcet,  ZoodogtaildirectionSiniscalchi, ZOOcatunderstandingbehaviorEllis}. It is important to note that the mapping between tail movements and emotional expressions may differ across species, and in some cases, even conflict across different species. For example, in pigs, a curled tail may indicate a positive emotional state, whereas this phenomenon is uncommon in cats and dogs~\cite{ZOOPigreviewCamerlink, ZOOPigtailemotionMarcet}. In dogs, tail inclination to the left or right can be associated with either positive or negative emotions~\cite{ZOOdogspecialRuge, ZoodogtaildirectionSiniscalchi}. For clarity and usability, we focused only on the most common and widely recognized cases as references. Certain animals with atypical tail morphology, such as sheep, whose tails are noticeably shorter than those of the other species included, were not represented in this table~\cite{ZOOsheepandcattleMachado}. 

We found that an upward tail position is typically associated with positive emotions, whereas a downward tail position generally indicates negative emotions~\cite{ZooCatzhao, ZOOallanimalKiley, ZOOdogspecialRuge, ZOOcatunderstandingbehaviorEllis}. Additionally, the amplitude and speed of tail movement further modulate the expression of emotional states. These emotions were originally defined by prior studies. We further identified their potential correspondences with the Ekman six basic emotions model~\cite{ekman1999basic} to facilitate the development of our mapping scheme. The specific emotions include happiness, Social Approach, Curiosity, Fear, Frustration, and Defensiveness. These emotions capture the majority of cases in which animals use their tails to express emotions. \autoref{tab:emotions} presents the mapping between tail movements and emotions. Each mapping is accompanied by a description of the corresponding animal behavior.

\subsection{Tail in Human-Robot Interaction - \autoref{tab:hri}}
HRI researchers primarily aim to convey emotions through tail movements that are intuitive to users, and a considerable body of work has emerged in this area~\cite{ROBOxietail, ROBOcatliketail, ROBOismarVRtail, ROBOtaildesign, ROBOsinghDogtailfullimp, ROBOsinghDogTailIntention, ROBOtailLeeStudleyIntention, ROBObioinspconveyemotion, ROBOemotionmodel}. However, our review revealed that different studies have employed varying parameterization standards to define tail actions. For example, Singh et al. used parameters such as speed, wag-size, height, wagging type (horizontal, vertical, circular), and postures (high, low, neutral) to represent emotions~\cite{ROBOsinghDogtailfullimp, ROBOsinghDogTailIntention}, while Wang et al. defined tail actions using frequency, amplitude, and waveform~\cite{ROBOcatliketail}. Furthermore, the emotion models employed also differ across studies: Macdonald et al. utilized a customized version of the Ekman model~\cite{ROBOismarVRtail}, whereas \citet{ROBOghafurianMiro} adopted a fully customized emotional framework.  Given these uncertainties, we adopted the Ekman emotion model, the most widely recognized and commonly used framework in HRI~\cite{ekman1999basic}. We also standardized the tail movement parameters, which will be detailed in the following section.



\subsection{Insights Merged from Both Aspects - \autoref{tab:my-table}}
We developed our mapping scheme based on insights from both HRI and zoology. In addressing discrepancies identified during our review, particularly regarding emotion models. We chose to adopt the Ekman six basic emotions model, as it is commonly employed in previous HRI research~\cite{ROBOismarVRtail, ROBOcatliketail}. For the parameterization of tail movements, we drew on the work of Singh et al., as their set of tail movement parameters is the most comprehensive and provides a detailed description of various action patterns~\cite{ROBOsinghDogtailfullimp, ROBOsinghDogTailIntention}. Specifically, we adopted three primary movement types: Continuous Wag (horizontal, vertical, circular), Action Gesture (raising, lowering), and Static Posture. The mapping between tail movements and the Ekman six basic emotions model is as follows: \textbf{Happy}, can be represented by three types of tail movements, \textit{Horizontal Wagging} (in the Continuous Wagging category), characterized by \textit{high speed, large wag size, and a high tail position (height)}~\cite{ROBOcatliketail, ROBOsinghDogtailfullimp, ROBOsinghDogTailIntention}. \textit{Circular Wagging}, performed at \textit{high speed}~\cite{ROBOsinghDogtailfullimp, ROBOsinghDogTailIntention}. \textit{Static Posture}, with the tail held in a \textit{high position}~\cite{ZOOallanimalKiley, ZoodogtaildirectionSiniscalchi}. \textbf{Angry} can be represented by two types of tail movements. \textit{Vertical Wagging} is performed at \textit{high speed} and with a \textit{large wag size}~\cite{ZoodogtaildirectionSiniscalchi, ZOOdogspecialRuge, ROBOsinghDogtailfullimp}. \textit{Circular Wagging}, performed at a \textit{medium speed}~\cite{ROBOsinghDogtailfullimp, ROBOsinghDogTailIntention}. \textbf{Sad} is represented by the \textit{Lowering action} in the \textit{Action Gestures} category~\cite{ZooCatzhao}, characterized by a \textit{slow movement} and a \textit{low tail position}. \textbf{Scared} can be represented in two ways: \textit{Horizontal Wagging} , performed at high speed with \textit{small wag size} and a \textit{low tail position} to indicate trembling~\cite{ZooCatzhao, ZOOcatunderstandingbehaviorEllis}, \textit{Static Posture}, with the tail held at a \textit{medium position}~\cite{ROBOsinghDogtailfullimp}. For detailed mappings, please refer to \autoref{tab:my-table}.





\section{Implementation}
\subsection{Physical Tail}
\begin{figure}[htbp]
    \centering
    \includegraphics[width=0.5\linewidth]{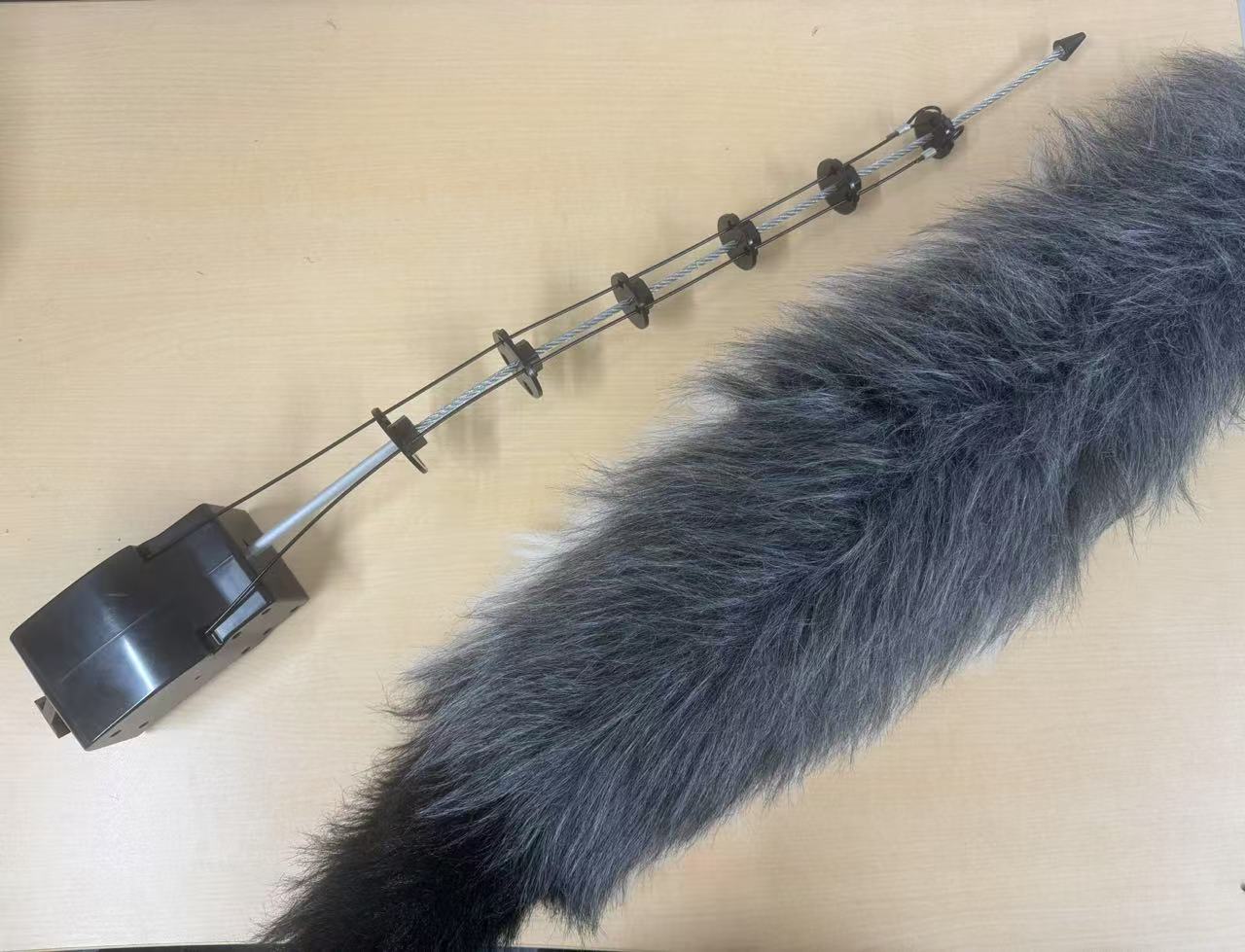}
    \caption{Implementation of our physical tail. continuum-type tail structure with 7 segments and fur.}
    \label{fig:tail}
\end{figure}

We chose the continuum-type tail structure, as it is currently one of the most common design~\cite{ROBOtailsurveysaab}. The tail prototype was custom-made by Tail Company. It consists of seven segments (with one as the tail tip) and is actuated by two 5521MG-equivalent 180° 20 kg digital servos, which drive two rigid wires to produce tail motions. Each servo is powered by a 103665 LiPo battery. The exterior of the tail is covered with synthetic fur approximating that of cats and dogs. This design enables the tail to perform basic horizontal and vertical wagging motions at various speeds. Additionally, the tail’s parameters can be adjusted and programmed via a dedicated app. \autoref{fig:tail} shows the overall shape and structure of the tail.

\subsection{Tail Motion}

\begin{figure*}[ht!]
    \centering
    \includegraphics[width=0.8\linewidth]{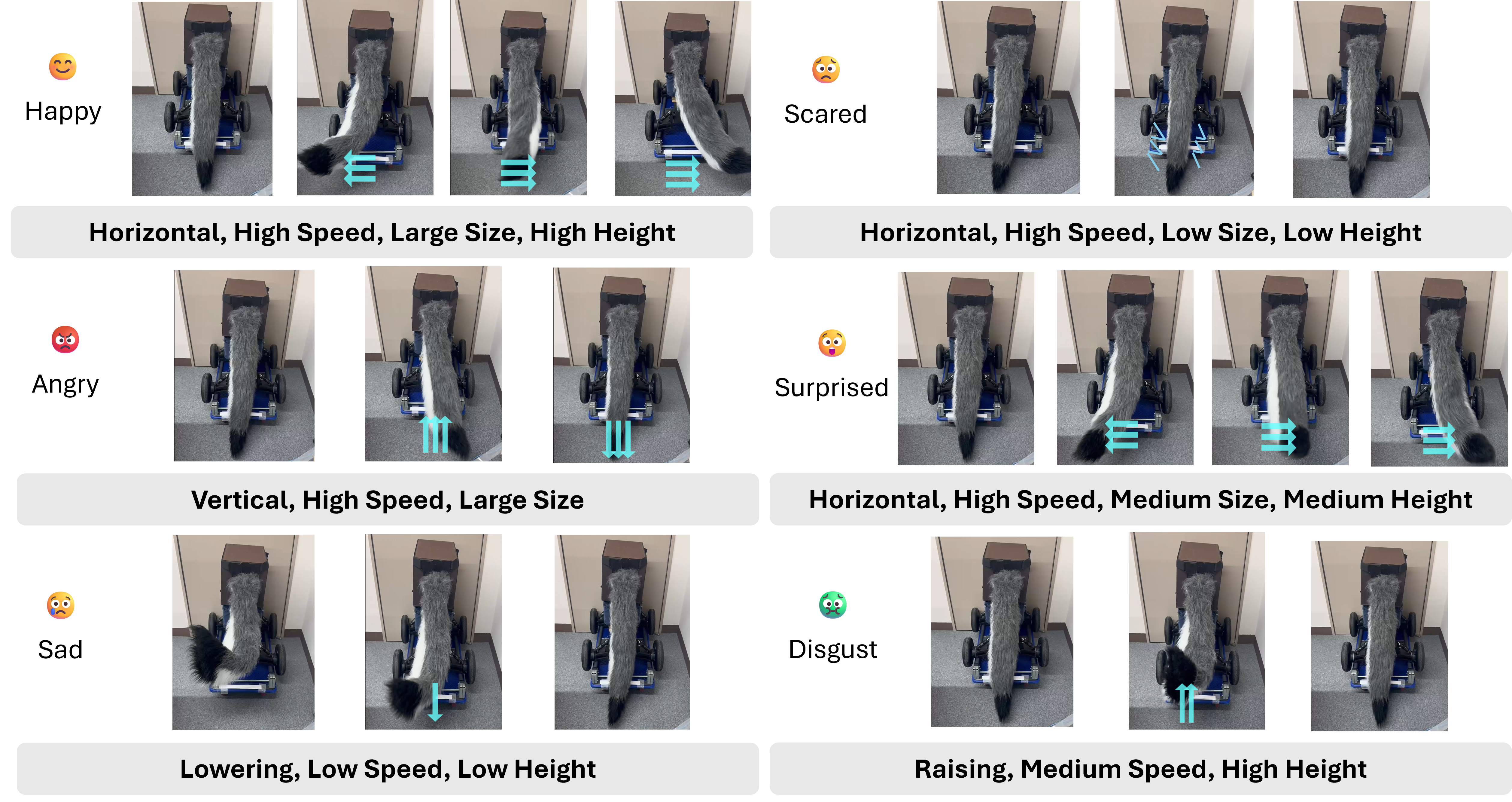}
    \caption{The tail motion we implemented to convey specific emotion by performing the designated sequence of motion parameters. The grey box shows the parameters, summarized in the mapping scheme, applied here.
    }
    \label{fig:scenarios-emotions}
\end{figure*}

We designed a mapping between emotions and tail movements. It is based on the functional characteristics of the tail and draws inspiration from our established mapping scheme. The specific motion descriptions are as follows: \textbf{Happy}: The tail swings \textit{horizontally} with a \textit{large wag size} at a \textit{high speed} in \textit{high height}. \textbf{Angry}: The tail wags \textit{vertically} with a \textit{large wag size} at a \textit{high speed}. \textbf{Sad}: The tail is \textit{lowering} at a \textit{low speed}. \textbf{Scared}: The tail exhibits a trembling motion, specifically in \textit{horizontal wagging}, \textit{high speed}, and maintains a \textit{low height}. \textbf{Surprised}: The tail is wagging horizontally at a high speed with medium wag size and medium height. \textbf{Disgust}: The tail is \textit{raising} to a \textit{high height} with \textit{medium speed}. A detailed breakdown of these actions is illustrated in \autoref{fig:scenarios-emotions}.

% ...

\section{User Study}

\subsection{Scenario}
Our user study was designed around four scenarios. Two of these scenarios focused on interactions involving Automated cars, while the other two centered on delivery robot interactions.

Currently, AV interactions can be broadly categorized into two types: those involving VRUs and non-VRUs~\cite{markVulnerable}. The first type typically refers to interactions between AVs and pedestrians. Pedestrians need to accurately interpret the AV’s intentions to cross the street safely. The second type involves interactions between drivers of conventional vehicles and AVs, where human drivers must quickly understand the AV’s behavior to prevent accidents.

For delivery robot scenarios, robots are generally small and physically vulnerable, requiring them to share urban spaces with other human road users. When navigating sidewalks, delivery robots may encounter unexpected situations in which they need to seek assistance from nearby pedestrians. Such help-seeking interactions are a frequently discussed topic in the field of AV and service robotics, and thus were also included as a key aspect of our evaluation.

\paragraph{Scenario 01: Weak Robot (WR) Interaction}
The participant takes the role of a hurried pedestrian. A delivery robot ahead needs assistance with pressing the elevator button. The introduction was:
\textit{"You are a passerby in a university building. On your way, you see a delivery robot telling you to assist with pressing the elevator button in front of the elevator. Please observe them and answer the questions."}

\paragraph{Scenario 02: Weak Robot Trapped}
The participant acts as a pedestrian. A delivery robot ahead is stuck and needs assistance to remove those objects. The introduction was:
\textit{"You are a passerby in a university building. On your way, you see a delivery robot stuck at the building entrance, and it tells you to remove the object that trapped it. Please observe them and answer the questions."}

\paragraph{Scenario 03: VRU Interaction}
The AV needs to reverse while the participant is positioned behind the car. The introduction was:
\textit{"An autonomous car is in front of you and is about to reverse. You are standing approximately 2 meters behind the vehicle. The vehicle is telling you that it is going to park here and needs you to move.  Please observe them and answer the questions. "}

\paragraph{Scenario 04: Non-VRU Interaction}
The participant assumes the AV is informing that it wants to change lanes. The introduction was:
\textit{"You are currently driving, and there is an autonomous car that wants to change your lane. He wants to give you this information. Please observe them and answer the questions."}

\subsection{Measurements}

Participants were asked to specify the exact emotion, including neutral. This was used to assess whether the tail movement evoked the emotion we defined.


We used the subscales \textit{Predictability / Understandability}  (\textit{Predictability}) and \textit{Trust} of the \textit{Trust in Automation} questionnaire~\cite{korber2018theoretical}.
\textbf{Predictability} is based on four statements (two direct: ``The system state was always clear to me.'', ``I was able to understand why things happened.''; two inverse: ``The system reacts unpredictably.'', ``It's difficult to identify what the system will do next.'') using 5-point Likert scales (1=\textit{Strongly disagree} to 5=\textit{Strongly agree}).
\textbf{Trust} is measured on the same 5-point Likert scale via two statements (``I trust the system.'' and ``I can rely on the system.''; higher is better).

One question using a 5-point Likert scale to collect participants' feelings regarding the safety when interacting with AVs ('How safe did you feel?').
We also used the subscales \textit{Emotions} of the \textit{Negative Attitude toward Robots} questionnaire~\cite{NegaAttituQuestionna}. \textit{Emotions} with 3 inverse statements using the 5 Likert scale mentioned above ('I would feel relaxed talking with autonomous vehicles.', 'If autonomous vehicles had emotions, I would be able to make friends with them.', 'I feel comforted being with autonomous vehicles that have emotions.'). 
One question using a 5-point Likert scale to collect participants' feelings regarding the safety when interacting with AVs was used ('How safe did you feel?') (1=\textit{Not at all} to 5=\textit{Totally}).
At the end of each scenario, participants could provide optional, open-ended feedback based on their recent experience. 

\subsection{Participants}
We recruited 21 participants (14 male, 7 female, 0 non-binary; M=25.92, SD=1.72 years old) through the university’s social media platforms. One is a bachelor student, 11 are Master students, three are employees, four are PhD students, one has no occupation, and one selected "other". 13 have had a pet, 11 of 13 have more than 1 year of experience. Each participant received a 2,000 yen Amazon gift card upon completion of the study.

\subsection{Procedure}
The user study was conducted online, with participants receiving the experiment link via email. Upon accessing the webpage, participants were presented with a written introduction outlining the study’s purpose and their responsibilities. After this introduction, they were asked to provide basic demographic information.

Each scenario was introduced with a textual description, and, after confirming participants’ understanding, a video corresponding to the condition was shown. Following each video, participants completed a set of survey questions. In total, participants experienced 28 trials (4 scenarios - called \Scenario, 2 AVs, and 2 delivery robot scenarios × 7 external cues - called \Emotion, including 6 tail motions and 1 text, text condition matched as neutral emotion), and participants viewed a total of 28 conditions, comprising 4 scenarios with 7 external cues each. For each scenario, one condition was text-only, while the remaining six conditions corresponded to tail movements mapped to our Ekman emotion model, with all videos and conditions randomized to mitigate order effects. The study took $\approx$ 60 minutes to complete.

\section{Results}

\subsection{Data Analysis}
Before every statistical test, we checked the necessary assumptions (e.g., normality of distribution).
For non-parametric data, we used ARTool~\cite{wobbrock2011art} and Holm correction for post-hoc tests (via Dunn's test). The procedure is abbreviated with ART.
R 4.5.0 and RStudio 2025.5.0 were employed. All packages were up-to-date in June 2025.

\subsection{Emotion Recognition Accuracy}

\begin{figure}[ht!]
    \centering
    \small
    \includegraphics[width=0.9\linewidth]{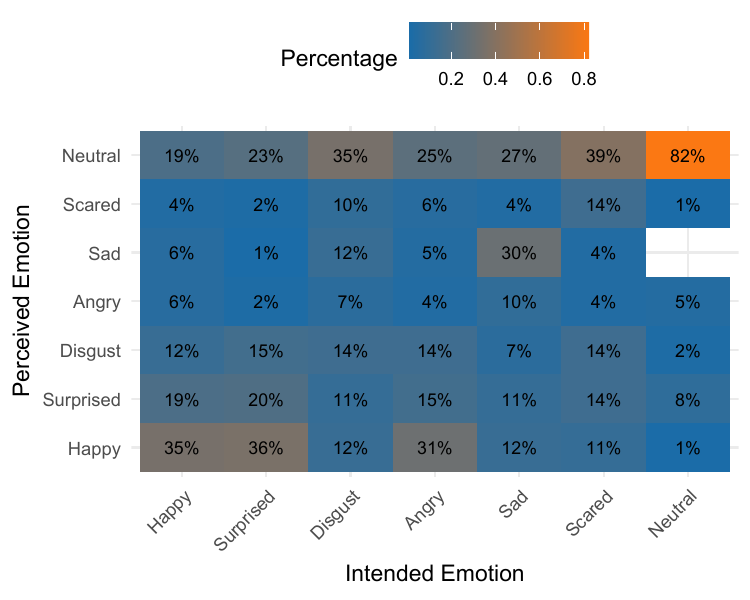}
    \caption{Confusion Matrix for perceived emotion.}\label{fig:conf}
    \Description{This figure shows}
\end{figure}

The participants can't recognize the tail motion emotion correctly in most cases. Out of the 588 trials, participants were only able to guess the correct emotion 167 times (28.4\%). \autoref{fig:conf} shows the confusion matrix for the different emotions. 


\begin{itemize}
\item The effect of \Emotion [Angry] is statistically significant and negative (beta = -2.58, 95\% CI [-4.79, -0.38], p = 0.022; Std. beta = -2.58, 95\% CI [-4.79, -0.38])
\item The effect of \Emotion [Neutral] is statistically significant and positive (beta = 2.85, 95\% CI [1.12, 4.59], p = 0.001; Std. beta = 2.85, 95\% CI [1.12, 4.59])
\end{itemize}

\subsection{Trust and Predictability}
The ART found a significant main effect of \Scenario on Trust (\F{3}{60}{4.95}, \p{0.004}). 
A post-hoc test found that Delivery robot: press button was significantly higher (\m{3.28}, \sd{1.01}) in terms of \Trust compared to Automated vehicle: change lane (\m{2.89}, \sd{1.08}; \padj{0.015}). A post-hoc test found that Delivery robot: trapped was significantly higher (\m{3.32}, \sd{0.90}) in terms of \Trust compared to Automated vehicle: change lane (\m{2.89}, \sd{1.08}; \padj{0.005}).

The ART found a significant main effect of \Emotion (\F{6}{120}{4.66}, \pminor{0.001}) and of \Scenario on Understanding (\F{3}{60}{4.87}, \p{0.004}). 
A post-hoc test found that Delivery robot: press button was significantly higher (\m{3.43}, \sd{0.75}) in terms of \Understanding compared to Automated vehicle: change lane (\m{3.06}, \sd{0.83}; \padj{0.003}). A post-hoc test found that Delivery robot: press button was significantly higher (\m{3.43}, \sd{0.75}) in terms of \Understanding compared to Automated vehicle: reverse (\m{3.10}, \sd{0.88}; \padj{0.002}). 
A post-hoc test found that Happy was significantly higher (\m{3.47}, \sd{0.83}) in terms of \Understanding compared to Sad (\m{3.01}, \sd{0.81}; \padj{0.015}). A post-hoc test found that Happy was significantly higher (\m{3.47}, \sd{0.83}) in terms of \Understanding compared to Scared (\m{3.02}, \sd{0.88}; \padj{0.006}).

\subsection{Negative Emotions}

The ART found a significant main effect of \Emotion on negative emotions towards the system (\F{6}{120}{2.54}, \p{0.024}) and of \Scenario on negative emotions towards the system (\F{3}{60}{2.80}, \p{0.047}).  Neither for \Scenario nor for \Emotion did post-hoc tests find any significant differences.

\subsection{Safety}
\begin{figure}[ht!]
    \centering
    \small
    \includegraphics[width=\linewidth]{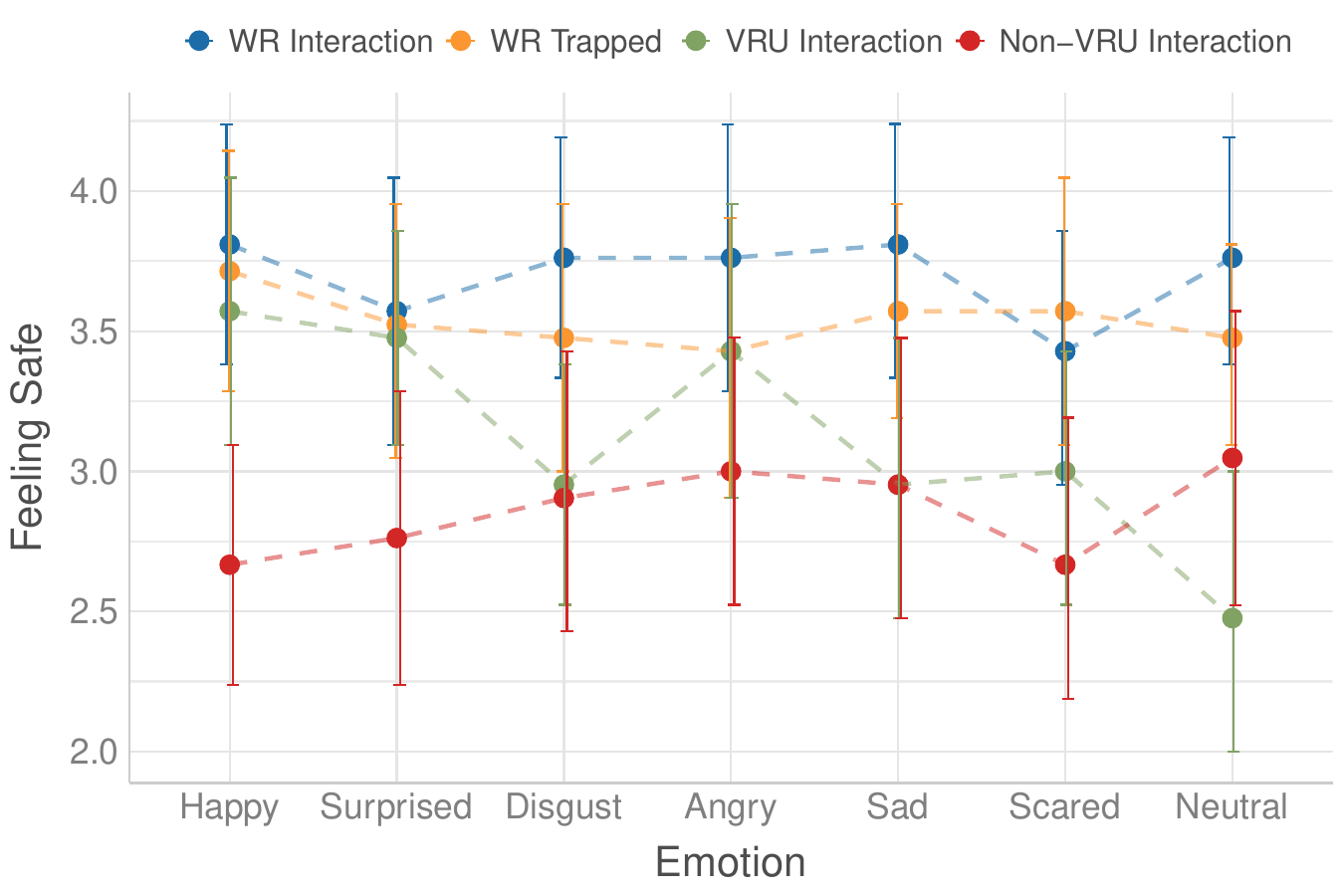}
    \caption{Interaction effect on feeling safe.}\label{fig:ie_safe}
    \Description{This figure shows}
\end{figure}

The ART found a significant main effect of \Scenario on Safe (\F{3}{60}{11.16}, \pminor{0.001}). The ART found a significant interaction effect of \Emotion $\times$ \Scenario on Safe (\F{18}{360}{1.76}, \p{0.029}; see \autoref{fig:ie_safe}). While WR interaction was almost always rated highest and Non-VRU lowest in terms of feeling safe, there was no distinguishable pattern for the other two scenarios.

\subsection{Open Feedback}
9 participants provided meaningful comments. Of these, 5 participants noted that when the tail’s movements (the emotion that conveyed) aligned with the scenario, it made them happier or influenced their decision-making. \textit{If an action matches the situation and the cues, it feels satisfying. But if they don’t align, it can backfire—just like when someone lies, but their body language gives it away. We sense the inconsistency and feel a sense of discomfort or even anger.} The other 4 participants indicated that the tail could be distracting, and that focusing solely on the textual information might be more effective in the absence of the tail. \textit{I actually rely more on text to judge the robot's meaning, and the meaning conveyed by the tail is very subtle.}

\section{Discussion}


\textbf{\textit{Using Tail to convey accurate emotions in AV context}}
We did not achieve high accuracy in emotion recognition. Although previous HRI research has shown that tail movements can effectively convey emotions and that users can perceive these expressions~\cite{ROBOsinghDogtailfullimp, ROBOsinghDogTailIntention, ROBOtailLeeStudleyIntention}, this finding did not hold in our study. We attribute this discrepancy to several factors. First, the outdoor environment in AV-pedestrian interaction is considerably more complex; participants must analyze the situational context, consider their own safety, and account for other road users. This complexity may impose a greater cognitive load, making it more difficult for participants to interpret the emotions conveyed by the tail. This added cognitive demand may be a key factor contributing to the lower recognition rates for tail-based emotional cues.

\textbf{\textit{Trust and Understanding with tail}}
Our results did not show that the introduction of a tail significantly enhanced users’ trust in AVs. This finding suggests that tail-based eHMI may be insufficient to foster trust in AVs within outdoor environments. The absence of a human driver means that animal-inspired elements alone may not increase trust, people may be hesitant to rely on AVs or delivery robots designed with animal-like features for critical tasks.

\textbf{\textit{Tail movements need to be contextually appropriate to the specific scenario}}
The use of a tail can help road users interpret the intentions of AVs, provided certain conditions are met. Specifically, the emotional expression conveyed by the tail must correspond appropriately to the scenario. \textit{“I didn't perceive the rest of the emotions on the tail. However, the 'happy' movements made me feel more comfortable and really more in the system. ”} Some participants reported that, when the tail movement and scenario were well aligned, the tail encouraged interaction and enhanced the positive experience. Conversely, when the tail movement was not contextually appropriate, the tail element could have a counterproductive effect, with some participants preferring to focus solely on text cues presented on the vehicle.

\textbf{\textit{Safety risks}}
The use of tail-based eHMI may pose potential safety risks. Many participants reported that, in certain scenarios, the tail excessively attracted their attention and required additional time to interpret its meaning. This suggests that tail-based cues may not be suitable for interactions involving VRUs, especially in situations such as street crossings, which inherently carry a higher risk of accidents. These findings highlight the importance of considering the risk profile and contextual appropriateness of tail-based interfaces in different AV scenarios, and motivate future research to investigate the suitability of tail cues across a range of interaction contexts.

\textbf{\textit{Limitation and Future Works}}
One limitation of this study is that we were unable to test all of the tail motions included in our proposed mapping scheme. It is possible that some tail movements not evaluated in this study may be more suitable for AV scenarios. Another limitation is that our study relied solely on subjective questionnaires. Incorporating objective measures, such as physiological data, to monitor participants’ responses while viewing AV scenarios would allow for a more comprehensive analysis of the interface’s effects.
In future research, it will be important to explore the key differences between AV scenarios and indoor pet robot contexts to further optimize the development of eHMIs. Differences in attitudes toward tail-based interfaces among various groups in different pet experiences also represent an important area for further investigation. Understanding how the safety requirements of different AV scenarios relate to the suitability of tail-based eHMI is a critical area for future investigation. It will be important to identify contexts in which the use of tail cues is both effective and appropriate.

\section{Conclusion}
This work proposed a tail-based eHMI, aimed at encouraging interaction between road users and AVs. We implemented this tail-based eHMI in AV-pedestrian scenarios and conducted a user study involving 21 participants. Our results indicate that tail-based interfaces in AV contexts may not convey accurate emotion, but can improve the interaction experience in AV contexts. Our research provides inspiration for the application of animal-inspired external interfaces in future HRI studies. In subsequent work, we will focus on identifying and optimizing scenarios where tail-based cues are most effective, with the aim of fostering safe and harmonious AV interaction environments.

\section*{Open Science}
All data and code are available under this GitHub repository: \url{https://github.com/Kulipajun/HAI-2025---Tail-Project}.

\begin{acks}
We thank all study participants.
This work was supported by a Canon Research Fellowship and by JST CRONOS, Grant Number JPMJCS24K8, Japan.
\end{acks}

\bibliographystyle{ACM-Reference-Format}
\bibliography{software}

\appendix

\end{document}